\documentclass[useAMS,usenatbib]{mn2e}

\usepackage[latin1]{inputenc}
\usepackage{graphicx}
\usepackage{epstopdf}

\usepackage{subfigure}
\usepackage{rotate}
\usepackage{amsmath, float}
\usepackage{txfonts}
\usepackage{fleqn}
\usepackage{color}
\usepackage{comment}

\usepackage{mathrsfs} 
\usepackage{booktabs}
\usepackage{colortbl}

\newcommand{\be}{\begin{equation}}
\newcommand{\ee}{\end{equation}}
\newcommand{\bdm}{\begin{displaymath}}
\newcommand{\edm}{\end{displaymath}}
\newcommand{\bea}{\begin{multline}}
\newcommand{\eea}{\end{multline}}
\newcommand{\ba}{\begin{align}}
\newcommand{\ea}{\end{align}}

\renewcommand{\sigma}{\varpi^2}

\newcommand{\apj}{ApJ}			
\newcommand{\apjl}{ApJLett}		
\newcommand{\aap}{A\&A}			
\newcommand{\mnras}{MNRAS}		
\newcommand{\prc}{Phys.~Rev.~C}		
\newcommand{\prd}{Phys.~Rev.~D}		

\def\simlt{\mathrel{\hbox{\rlap{\hbox{\lower4pt\hbox{$\sim$}}}\hbox{$<$}}}}
\def\simgt{\mathrel{\hbox{\rlap{\hbox{\lower4pt\hbox{$\sim$}}}\hbox{$>$}}}}

\title[Quark deconfinement in the proto-magnetar model of LGRB]
{Quark deconfinement in the proto-magnetar model of Long Gamma-Ray Bursts}
\author[A.~G. Pili, N. Bucciantini, A. Drago, G. Pagliara, L. Del Zanna ]{
A.~G. Pili$^{1,2}$\thanks{E-mail: pili@arcetri.astro.it},
N. Bucciantini$^{2,1}$,
A. Drago$^{3}$,
G. Pagliara$^{3}$,
L. Del Zanna$^{1,2}$\\
$^{1}$Dipartimento di Fisica e Astronomia, Universit\`a degli Studi di
Firenze and INFN Sez. di Firenze, Via G. Sansone 1, I-50019 Sesto F.~no  (Firenze), Italy\\ 
$^{2}$INAF - Osservatorio Astrofisico di Arcetri, Largo E. Fermi 5, I-50125 Firenze, Italy\\
$^{3}$Dipartimento di Fisica e Scienze della Terra, Universit\`a di Ferrara and INFN Sez. di Ferrara, Via Saragat 1, I-44122 Ferrara, Italy}

\begin{document}
 
\date{Accepted / Received}

\maketitle

\label{firstpage}

\begin{abstract}

We investigate the possible implications of quark deconfinement on the phenomenology of Long Gamma-Ray Bursts
focusing, in particular, on the possibility to describe multiple prompt emission phases in the context of the 
\textit{proto-magnetar model}. Starting from numerical models of rotating Hadron Stars and Quark Stars in full general relativity
we track the electromagnetic spin-down evolution in  both the hadronic and quark phase, linking the two families through conservation of
baryon number and angular momentum. We give estimates of the timescales and the energetics involved in the spin-down process 
deriving, in the relevant spin range, the relation between the initial and the final masses and rotational energies, whenever hadron-quark
conversion is possible. We show how the results can be used in relevant astrophysical cases such as the double burst GRB 110709B. 

\end{abstract}

\begin{keywords}
gamma-ray bursts - star: magnetar  - star: neutron - dense matter -  gravitation
\end{keywords}

\section{Introduction}

The nature of the inner engine of Long Gamma-Ray Bursts (LGRBs) is
still one of the most interesting and unsolved problems in
astrophysics. While there is a compelling evidence that these events
are associated with the core collapse of massive stars, it is not yet
established whether the burst is produced by a disk
accreting onto a black hole (within the so-called Collapsar model,
 \citealt{Woosley93a}) or by the outflow emerging from a fast rotating
and highly magnetized proto-neutron star (within the so-called
proto-magnetar model, \citealt{Metzger_Giannios+11a}). It is also possible
that both these systems contribute in producing LGRBs, depending
on the initial conditions of the progenitor (in particular its mass,
spin frequency and magnetic field). The SWIFT discovery of a late time
activity, lasting up to $10^4$ s and present in a sizable
fraction of LGRBs, is more naturally interpreted within the
proto-magnetar model as due to the pulsar-like energy injection powered 
by the residual rotational energy left after the prompt emission
\citep{Dall'Osso:2010ah}.

Within the proto-magnetar model, an important ingredient which
regulates the temporal evolution of the jet and its gamma-ray luminosity
is the neutrino signal released by the star due to deleptonization
(the gradual neutronization of matter) and cooling. In particular,
neutrinos ablate baryonic matter from the surface and provide a tiny
amount of baryonic load which is crucial for an efficient internal
dissipation of the kinetic energy of the jet into gamma emission. When
the neutrino luminosity drops below $\sim 10^{50}$ erg/s (after a few
tens of seconds), the magnetization (or the Lorentz factor) is too
large and the prompt emission ends. It is not easy within the
proto-magnetar model to reactivate the inner engine and therefore to
describe multi-episodes of the prompt emission of LGRBs (e.g. \citealt{Zhang_Burrows+12a})  
or X-ray flares occurring during the afterglow
\citep{Zhang:2005fa}.

The proto-magnetar model has been developed assuming that the newly
born Compact Star (CS) is and remains a nucleonic star.  Motivated by
the measurements of very massive CSs
\citep{Demorest:2010bx,Antoniadis:2013pzd} and the hints of the
existence of very compact stellar objects (with radii close to 10km,
\citealt{Guillot:2013wu,Ozel:2015fia}), a two-families scenario has been
recently proposed where both Hadronic Stars (HSs) and Quark Stars
(QSs) exist in nature \citep{Drago_Lavagno+14a,Drago_Lavagno+15b,Drago_Pagliara15b}.
Within this picture, a conversion of a HS into a QS can take
place. In this paper we will discuss how this transition modifies the
proto-magnetar model, revitalizing the inner engine.  In particular,
we will investigate the following scenario: i) a proto-magnetar is
formed after a successful supernova explosion and its spinning down,
which is responsible for the emission of a LGRB, leads to a gradual
increase of the central density; ii) the increase of the density
allows the formation of heavy hadrons such as Delta resonances and 
hyperons; iii) once a critical amount of strangeness is formed through hyperons at a density
$\rho_{\rm crit}$, the HS converts (on a time-scale $\lesssim10$~s) into a QS. 
This picture has been qualitatively discussed in \citet{Drago_Lavagno+14a} 
for the case of non-rotating stars. Here we will
improve that work by including rotation and spin-down evolution along the line of
\citealt{Haensel_Bejger+2016}, and Refs. therein. 
Since both the initial hadronic and the final quark configurations are fast
rotating stars, we compute their structure as rigid
rotating stars beyond the Hartle approximation \citep{Hartle67a} through the XNS code,
which  adopts the conformally-flat approximation for the space-time metric 
(see \citealt{Bucciantini_Del-Zanna11a,Pili_Bucciantini+14a,Pili_Bucciantini+15a,Bucciantini_Pili+15a}).

A distinctive feature of the two-families model is that the
formation of a QS is accompanied by a rather large amount of energy
released in the conversion, of the order of $10^{53}$ erg. Moreover,
we will show that since the final QS configuration has a larger radius than 
the initial HS configuration, the conversion is accompanied by a
significant increase of the momentum of inertia with a corresponding
decrease of the rotational frequency.  We will discuss possible
phenomenological implications of the two-families scenario for the
light curves of LGRBs in connection with late time activity.
For the case of  QSs there are two possible ways to produce a jet
with the appropriate Lorentz factor. The first mechanism is based on
the ablation of baryonic material from the surface of the star as long
as the conversion front has not yet reached the surface \citep{Drago_Pagliara15b}. 
 A second mechanism is based on the rather large emission of electron-positron
pairs from the surface of the bare quark star \citep{Usov:1997ff,Usov:2001sw,Page:2002bj}. 
Both mechanism can be at work to produce late time activities.

In section~\ref{sec:evolution} we describe our model and present our numerical results.
In section~\ref{sec:discussion} we conclude, discussing the phenomenological implications.

\section{The quasi-stationary evolution of Compact Stars}

\label{sec:evolution}

In order to investigate the possible implications of  the two-families scenario for the phenomenology of GRBs we have computed a large
set of rigidly (rotation rate $\Omega=\rm{const}$) rotating equilibrium
models for both HSs and QSs.  
Given the typical cooling evolution of
newly formed proto-neutron stars \citep{Pons_Reddy+99a}, the assumption
of a cold star is justified at times larger than $\gtrsim 10$~s
after formation (the radius is relaxed to its final value).  These
numerical models are then linked together to describe the
quasi-stationary evolution of CSs under the effect of the magnetic
braking.  With this simple approach we can give a preliminary
estimate of both the timescales and the energetics involved in the
spin-down before and after quark deconfinement, which we assume to happen when the central baryon
density $\rho_c$ reaches a critical value of the order of $10^{15}  \,\mbox{g/cm}^3$ \citep{Drago_Lavagno+14a}.
Here we set $\rho_{\rm crit}=1.34\times 10^{15} \,\mbox{g/cm}^3$.
We also derive the range of initial masses for which a delayed HS to QS conversion
is possible.

Numerical models have been obtained using the equations of state (EoSs)
discussed in \citet{Drago_Lavagno+15b}. The stars have been discretized with
$\sim 500$ grid points in both the radial and angular direction and Einstein 
equations are solved with a semi-spectral method using 20 spherical harmonics.  
Since the typical
surface magnetic field invoked by the magnetar model for GRBs are of
the order of $10^{15}$ G our models are computed neglecting magnetic
field. Indeed, as discussed in \citet{Pili_Bucciantini+15a}, only
central magnetic field strengths higher than $\sim 10^{16}$ G,
corresponding to surface magnetic fields stronger than a few $10^{15}$
G, are able to modify the stellar global properties to a level
appreciable with respect to the overall accuracy of the numerical
scheme ( $\lesssim 10^{-3}$). Hence we can safely assume that the
stellar structure and the associated global quantities such as the
gravitational mass $M$, the baryonic mass $M_0$ (a proxy for the total
baryon number), the circumferential radius $R$ and the Komar angular
momentum $J$ (for definitions see \citealt{Pili_Bucciantini+14a} and
Refs. therein) do not depend on the magnetic field strength.

In the millisecond magnetar model for GRBs the typical rotation periods
invoked in the literature \citep{Metzger_Giannios+11a,Bucciantini_Metzger+12a} 
can be as high as $\sim 1$ ms. The maximum rotation rate of compact stars at the end 
of deleptonization has been investigated in several papers 
\citep{Goussard:1997bn,Villain:2003ey,Camelio:2016fan}. 
Its value depends on the initial physical conditions of the proto-neutron star 
(e.g. the entropy profile and a possible differential rotation) 
and on the compactness of the final cold and deleptonized configuration. If one adopts rather stiff 
EoS, the maximum rotational frequency ranges between $300$ and $600$ Hz
\citep{Goussard:1997bn,Villain:2003ey,Camelio:2016fan}. In our
analysis we use a rather soft hadronic EoS due to the formation of Delta resonances and 
hyperons and the maximum rotation frequency is expected to be larger.
In the following we limit the rotation of our HS models to a
maximum frequency $f=10^3$ Hz. On the other hand, for those 
QSs resulting from the conversion of a HS it is sufficient to limit the spin frequency to $600$
Hz.

\begin{figure*}
	\centering
	\includegraphics[width=.46\textwidth]{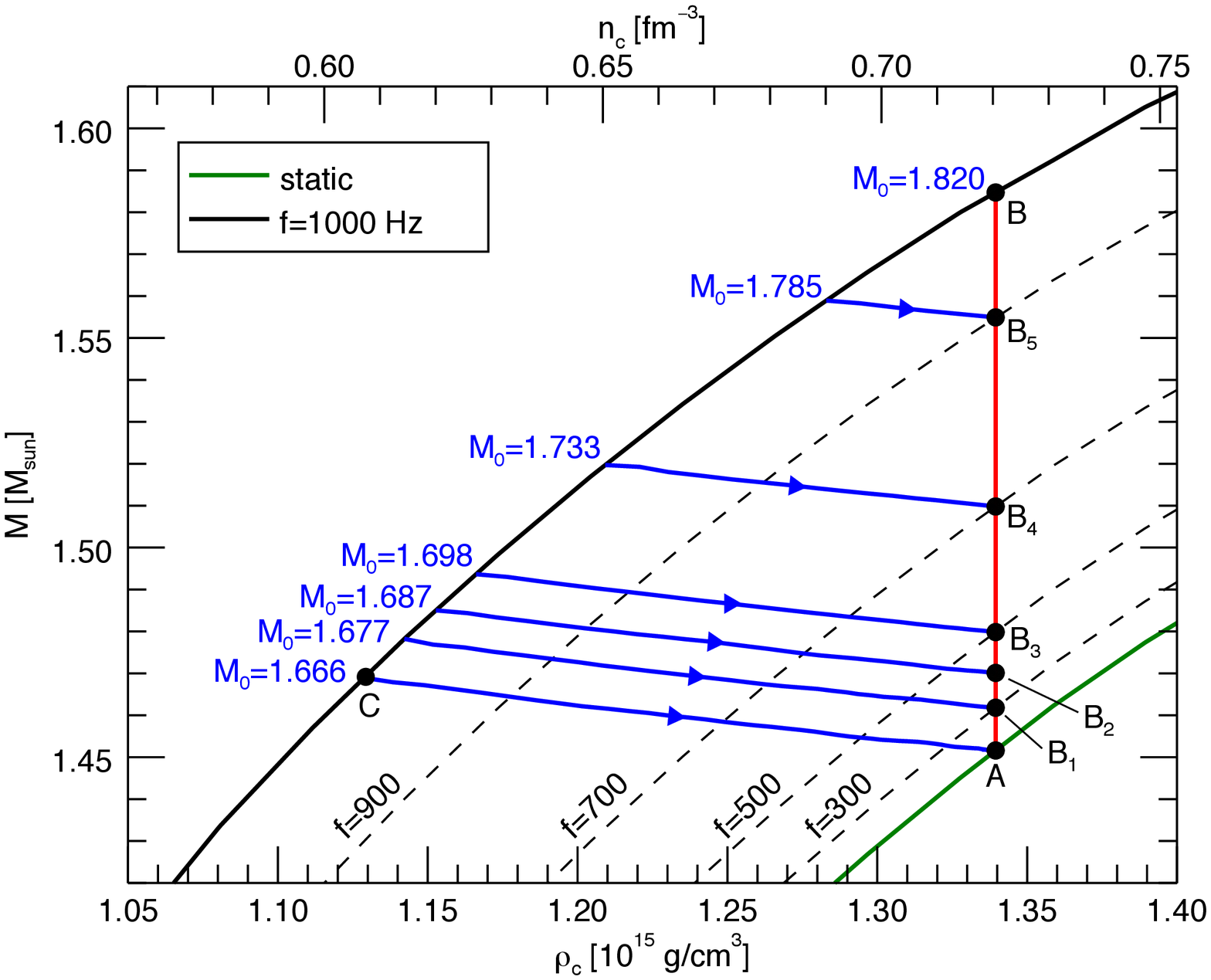}
	\includegraphics[width=.46\textwidth]{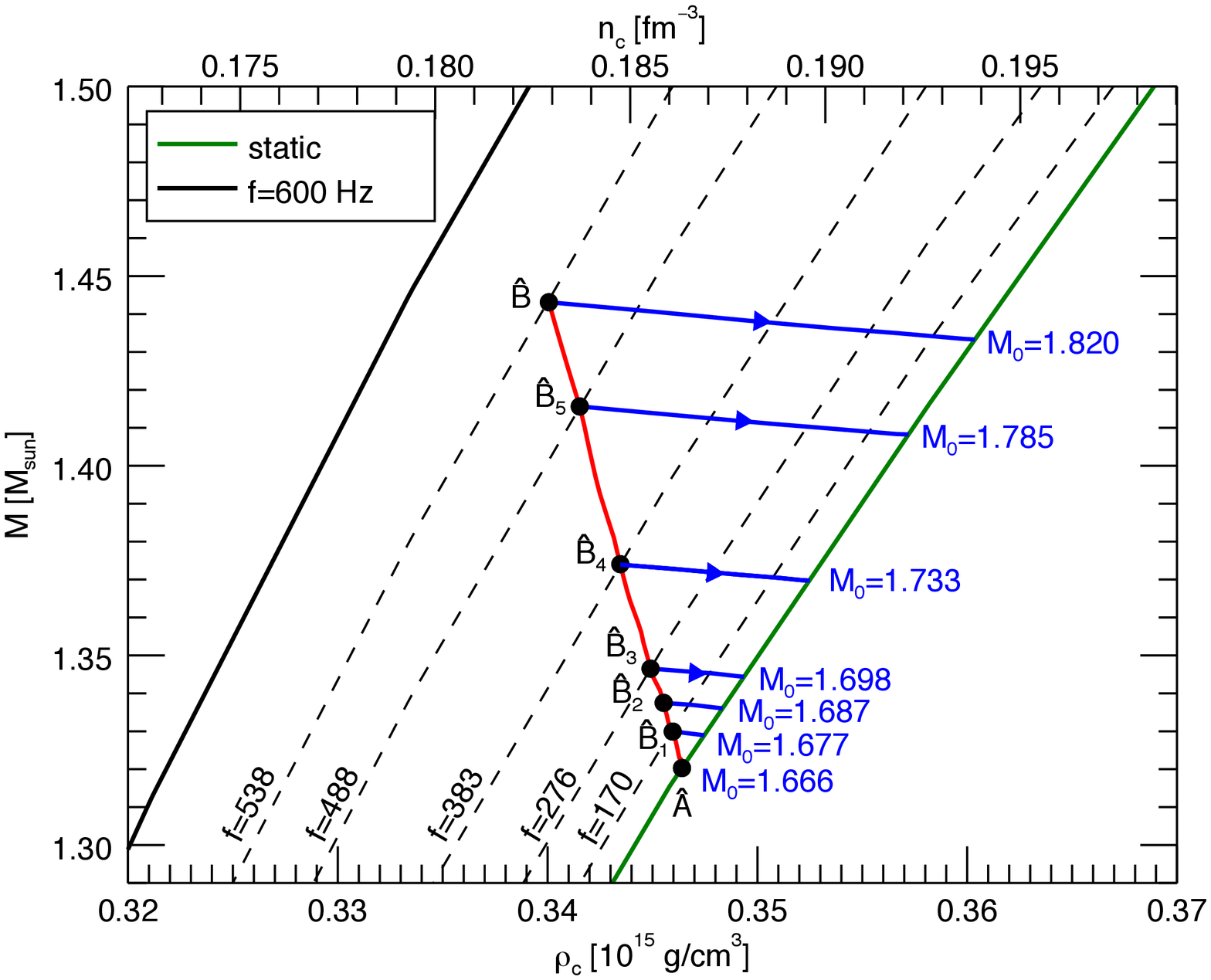} 
	\caption{Gravitational mass against baryon central density for sequences of HSs (left panel)
	and QSs (right panel) with constant spin frequency. In both cases 
	the green solid lines represent the non-rotating sequences while the black solid lines are
	sequences rotating at the maximum spin frequency considered in this work:
	 $1000 Hz$ for HSs and $600 Hz$ for QSs. Dashed black lines are sequences with intermediate values of $f$ while
	blue lines are equilibrium sequences at constant baryon mass $M_0$. The red solid lines locates HSs with 
	$\rho_{\rm c}=\rho_{\rm crit}$ while the red dashed line locates QSs originating from the conversion
	of HSs lying on the red solid line. Detailed results on the configurations labeled A 
	and Bs are listed in table~\ref{tab:transition}.
	 }
	\label{fig:Mrhoc.eps}
\end{figure*}

\begin{figure}
	\centering
	\includegraphics[width=.46\textwidth]{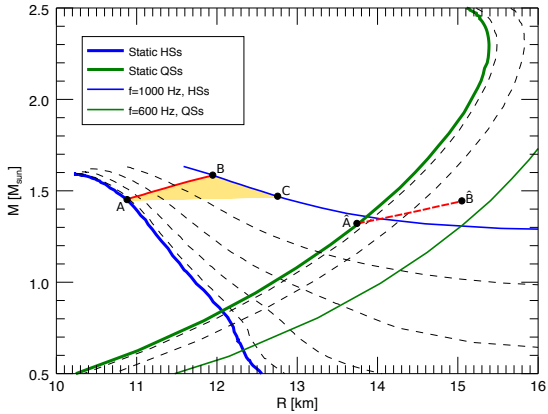} 
	\caption{ Gravitational mass as a function of the circumferential radius
          for both HSs and QSs.  Thin dashed lines are sequences of
          stars at a fixed frequency from the non-rotating
          configurations (thick solid blue and green lines) to the
          configurations rotating at the maximum frequency (thin solid
          blue and green lines) and spaced by $200$ Hz.
          The yellow region shows hadronic configurations centrifugally supported 
          against deconfinement. Red lines and labels are the same as in figure~\ref{fig:Mrhoc.eps}. 
              }
	\label{fig:MassRadREl.eps}
\end{figure}

The obtained equilibrium models are shown in figure~\ref{fig:Mrhoc.eps} in terms of central baryon
density $\rho_{\rm c}$ and gravitational mass $M$ for both HSs (left panel) and QSs (right panel). 
Here blue lines are sequences of equilibria for given baryonic mass and therefore they represent the
evolutionary path of HSs and QSs undergoing spin-down.  The same configurations are also shown 
in figure~\ref{fig:MassRadREl.eps} where we plot the mass-radius relation for different sequences of CSs at
constant spin frequency. There are two limiting configurations: the non-rotating configuration A
with $M({\rm A})=1.45 M_\odot$, $M_0({\rm A})=1.67 M_\odot$ and $R=10.9$ km; 
the 1 ms rotating configuration B with $M({\rm B})=1.58 M_\odot$,
$M_0({\rm B})=1.82 M_\odot$ and $R=12$ km. These define the mass range of
interest for delayed quark deconfinement.  The value of the critical
mass of the configuration A cannot be determined with high precision
not even in the scheme of the two families. The main uncertainties are
due to the dynamics regulating the appearance of baryonic resonances
and on the estimate of the nucleation time of the first droplet of
quark matter \citep{Bombaci:2016xuj}.

Massive HSs, that after their initial cooling (lasting for at most 
$\sim 10$~s after core collapse) end above the red line in
figure~\ref{fig:MassRadREl.eps}, have  $\rho_{\rm c} \geq \rho_{\rm crit}$ and they 
will decay immediately to a QS. On the other hand
those HSs having $M_0< M_0({\rm A})$ will never experience this
transition\footnote{For simplicity, we are not considering here the
process of nucleation of quark matter \citep{Bombaci:2016xuj}. In a
more realistic case the red lines connecting the configurations A
and B would transform in a strip whose width is connected to the
nucleation time.}. Since for isolated CSs the baryonic mass is a
conserved quantity, only those HSs having baryonic mass between
$M_0({\rm A})$ and $M_0({\rm B})$ can migrate into the QS branch when
their baryon central density rises to $\rho_{\rm{crit}}$ as a
consequence of spin-down.  The yellow shaded region of
figure~\ref{fig:MassRadREl.eps} shows these configurations.  All the
configurations having the same baryonic mass as A (for example the 1
ms rotator configuration C) take an infinite time to deconfine
(deconfinement takes place at zero frequency).  
Notice that along the evolutionary paths the variation of the
gravitational mass, and hence of the total energy, is at most of the
order of $\sim 1$\%.

As soon as the central density reaches the critical value the HS
decays into a QS in $\lesssim 10$~s.  Given the typical spin-down
timescales we assume that this transition is instantaneous and both
the baryon number (baryonic mass) and the angular momentum $J$ are
conserved.  The red dashed line in the right panel of figure~\ref{fig:Mrhoc.eps}
and in  figure~\ref{fig:MassRadREl.eps}  maps upon the QS
sequences the configurations which originate from quark deconfinement
of HSs.  With reference to our limiting configurations, the HS labeled
with A migrates to the configuration ${\rm \hat{A}}$ with
gravitational mass $M({\rm \hat{A}})=1.32 M_\odot$ , while the HS B
migrates to the configuration ${\rm\hat{B}}$ with
$M({\rm\hat{B}})=1.44M_\odot$. In both cases the total energy
released in the transition is of the order of $ \sim 0.1 M_\odot $.
Finally, after the formation, QSs can spin-down following the
evolutionary path of constant $M_0$ shown in the right panel of
figure~\ref{fig:Mrhoc.eps}. Detailed results are presented in
table~\ref{tab:transition} where we list the variation of the spin
frequency $\Delta f_{\rm d} $, of the gravitational mass $\Delta
M_{\rm d}$ and of the rotational kinetic energy $\Delta K_{\rm d}$
(where $ K:=\frac{1}{2} J \Omega$ ) during the phase transition for a
set of selected models. Since QSs have larger radii and hence also a
larger moment of inertia, conservation of the angular momentum implies
a drop in spin frequency. Interestingly we found that the variation of
the stellar radius is slightly larger ($\sim 10$\% between cases A and
B) at higher $M_0$. Hence transitions at higher $M_0$ are
characterized by a bigger percent variation of the spin frequency and
of the rotational energy. $\Delta K_{\rm d}$ only accounts for the
variation of the kinetic energy while the total change of energy is
given by the variation of the gravitational mass $\Delta M_d$ and it
will produce a reheating of the star.  In the range of interest,
$\Delta M_d \simeq 10^{53}$~erg and $\Delta K_d \simeq 10^{51-
  52}$~erg.  This significant amount of energy released by quark
deconfinement can be associated with a late time activity as we will
discuss in the following.
\begin{table}
\centering
\caption{ \label{tab:transition}
Global quantities at the phase transition for the configurations labeled in figure~\ref{fig:Mrhoc.eps}:
gravitational mass $M$ and spin frequency $f$ of the hadronic configuration at deconfinement;
difference between the spin frequency  $\Delta f_{\rm d}$, the gravitational mass $\Delta M_{\rm d}$, 
and the rotational energy  $\Delta K_{\rm d}$ before and after deconfinement.
}
\begin{tabular}{l*{6}{c}}
\toprule
\toprule
Label &$M_0$          & $ f$ &  $\Delta f_{\rm d}$   &   $M$             & $\Delta M_{\rm d}$    &  $\Delta K_{\rm d}$   \\
          &[$M_{\odot}$] &   [Hz]  &    [Hz]              &   [$M_{\odot}$]   & [$M_{\odot}$]  &   [$10^{52}$ erg]  \\
\midrule
 A                  &  1.666  &  0.00  &  0.00  &  1.452  &  0.132  &  0.00 \\
$\rm{B}_1$   &  1.677  &  300   &  130   &  1.462  &  0.132  &  0.20  \\
$\rm{B}_2$   &  1.687  &  400   &  177   &  1.470  &  0.133  &  0.35  \\
$\rm{B}_3$   &  1.698  &  500   &  224   &  1.480  &  0.133  &  0.57  \\
$\rm{B}_4$   &  1.733  &  700   &  317   &  1.520  &  0.136  &  1.18  \\
$\rm{B}_5$   &  1.785  &  900   &  413   &  1.555  &  0.139  &  2.14  \\
 B                  &  1.820  & 1000 &   462   &  1.585  &  0.142  &  2.80  \\
\bottomrule 
\end{tabular}
\end{table}

\begin{figure*}
	\centering
	\includegraphics[width=.46\textwidth]{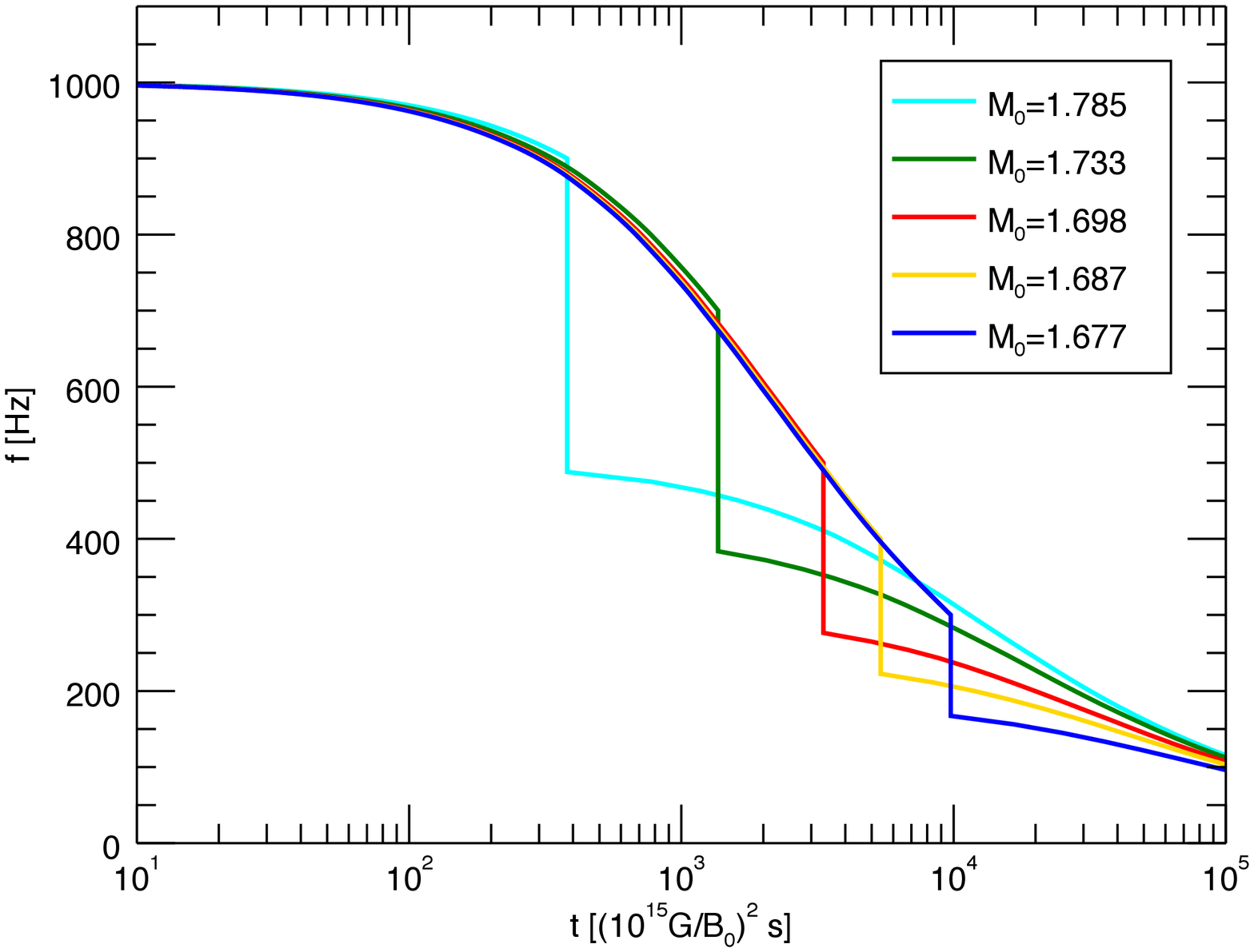}
	\includegraphics[width=.46\textwidth]{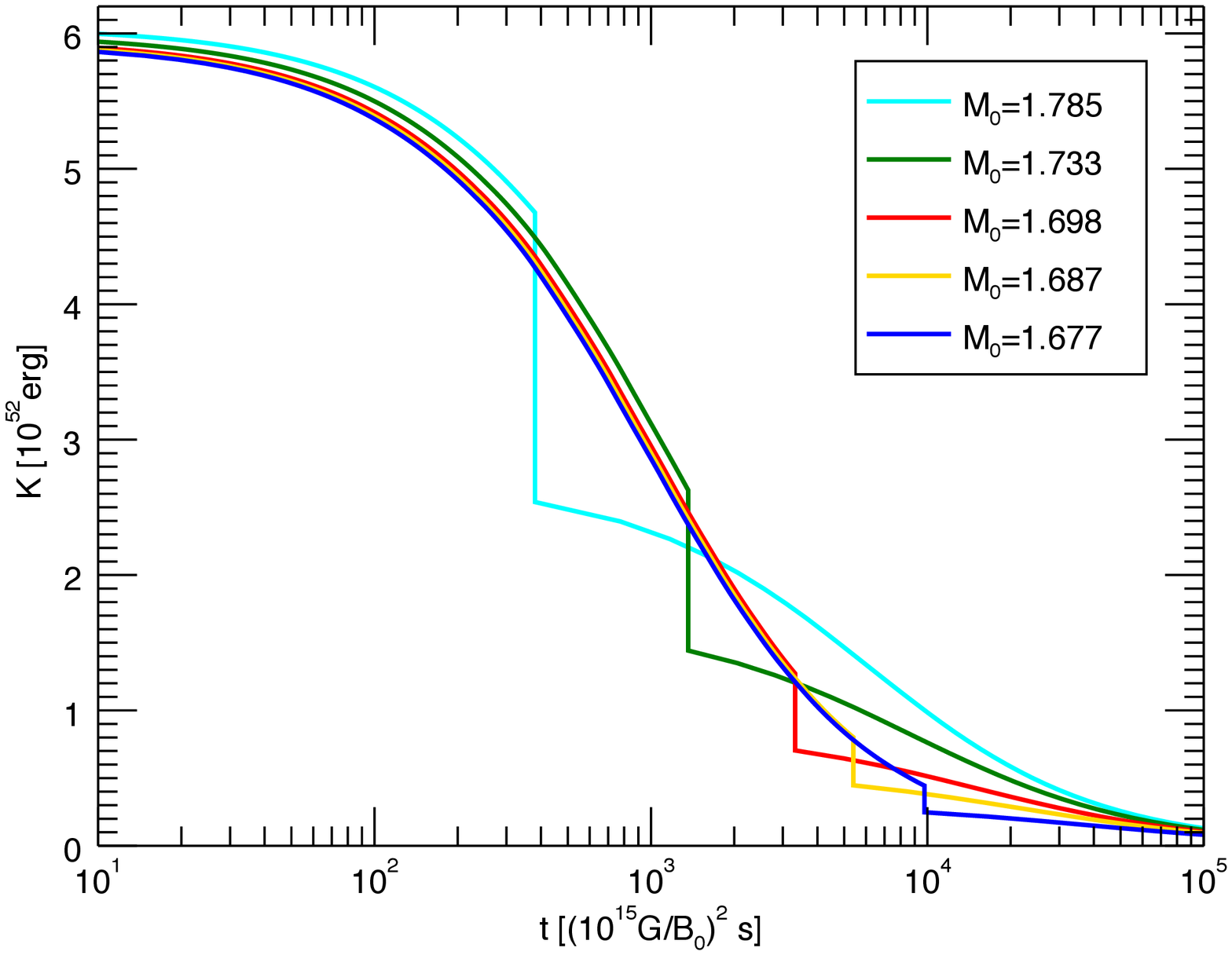} 
	\caption{ Time evolution of the spin-frequency (left panel) and of the rotational energy (right panel) for dipole
	magnetic field losses along sequences of constant baryonic mass $M_0$. Details about these sequences are listed
	in tables~\ref{tab:transition} and \ref{tab:denergy}.
              }
	\label{fig:tevolution}
\end{figure*}

In the magnetar GRB model the spin-down evolution is governed by magnetic torques. Therefore
in order to evaluate the evolutionary timescales of our models we solve, for simplicity, the spin-down 
formula for an aligned dipole rotator \citep{Spitkovsky06a}:
 \begin{equation}
\frac{ {\rm d} J }{{\rm d} t}=-\frac{B^2 R^6 \Omega^3}{4} ,
\label{eq:spind}
\end{equation}
where $B$ is the magnetic field strength at the pole and we assume
that the magnetic flux $\Phi= B R^2$ is constant during the
evolution. This choice gives only an upper limit to the timescale for
deconfinement: oblique rotators or mass loaded winds can have higher
torques \citep{Metzger_Giannios+11a} that can lead to faster spin-down
(up to a factor 10) in the first tens of seconds. Furthermore we
neglect the possibility of late time mass accretion
\citep{Bernardini_Campana+14a}, because it introduces extra degree of
freedom that cannot be easily constrained.  Moreover we also neglect
gravitational waves emission which is relevant only if the star owns a
very strong internal toroidal field larger than $\sim10^{16}$~G
\citep{DallOsso_Shore+09a}. In the mass range of interest we have
evaluated the spin-down timescale up to deconfinement, which is shown
in figure~\ref{fig:tevolution} and table~\ref{tab:denergy}, for each
equilibrium sequence of constant $M_0$ , assuming an initial surface
magnetic field $B_0=10^{15}$~G. Note that the spin-down age scales with
$B_0^{-2}$.  Conservation of the magnetic flux $\Phi$ is also assumed
to hold during the phase transition. The resulting magnetic field
strength is of the order of $0.5 B_0$ due to the change of the stellar
radius.
\begin{table}
\centering
\caption{ 
\label{tab:denergy}
Spin-down timescales to start quark deconfinement $\Delta t_{\rm sd}$ together with the associated variation of
the rotational kinetic energy $ \Delta K_{\rm sd}$  starting from an initial spin period $P_i$ for the equilibrium 
sequences shown in figure~\ref{fig:tevolution}.
We also report the spin-down timescales $\Delta t_{\rm q}$ (defined as the time needed to half the rotational frequency of the QS) 
and the corresponding rotational energy loss $\Delta K_{\rm q}$ 
after quark deconfinement. The initial magnetic field is of $10^{15}$ G.}
\begin{tabular}{l*{6}{c}}
\toprule
\toprule
$M_0$          & $P_i \rightarrow P_{\rm d}$ &  $\Delta t_{\rm sd}$   & $ \Delta K_{\rm sd}$ &  $\Delta t_{\rm q}$  &  $ \Delta K_{\rm q}$  \\
$[M_{\odot}]$ &   [ms]   &  &   [$10^{52}$ erg] &     & [$10^{52}$erg] \\
\midrule
1.666 & 1.0 $\rightarrow \infty$ & $\infty$& 5.91   & - & - \\
1.677 & 1.0 $\rightarrow$ 3.3	 &  2.7 hr  & 5.48  &  37 hr  & 0.19 \\
		 & 2.0 $\rightarrow$ 3.3    &  1.8 hr  & 0.82  &            &         \\
		 & 3.0 $\rightarrow$ 3.3    &  37 min  & 0.13  &            &         \\
1.687 &	1.0 $\rightarrow$ 2.5	  &  1.5 hr  & 5.13  & 21 hr  & 0.33  \\
         & 2.0 $\rightarrow$ 2.5    &  36 min  & 0.46   &            &         \\
1.698 & 1.0 $\rightarrow$ 2.0    &  55 min  & 4.68  &  14 hr & 0.53  \\
1.733 & 1.0 $\rightarrow$ 1.4    &  23 min & 3.37 &  8.2  hr & 1.20  \\		 
1.785 & 1.0 $\rightarrow$ 1.1    &  6 min & 1.37 &  5.4 hr & 1.95   \\	
1.820 & 1.0 $\rightarrow$ 1.0    &      0     &   0    &  4.6 hr &  2.41  \\
\bottomrule 
\end{tabular}
\end{table}

The time evolution of the spin frequency $f$ and of the rotational
energy $K$ is shown in figure~\ref{fig:tevolution} for the HS configurations labelled $B_1 - B_5$.  It is evident
that HSs with higher mass have a shorter lifetime before
deconfinement, since they reach the critical density at higher spin
frequencies. In table~\ref{tab:denergy} we list the time it takes to
start deconfinement and the associated rotational spin-down energy loss
$\Delta K_{ \rm sd}$ for different values for the initial spin period.

\section{Discussion and conclusions}
\label{sec:discussion}
Let us consider now the phenomenological implications of our scenario
for the evolution of proto-magnetars. The variation of the rotational
energy $\Delta K_{ \rm sd}$ displayed in table~\ref{tab:denergy} gives an 
estimate of the energy reservoir available to the HS before deconfinement which
occurs after a spin-down time scale $\Delta t_{\rm sd}$ that ranges
from several minutes to hours. Comparing $\Delta K_{ \rm sd}$ with the
typical energetics of the millisecond magnetar model for GRBs (figure
19 of \citealt{Metzger_Giannios+11a}) one notices that the values shown
in table~\ref{tab:denergy} are compatible with the requirement for
classical GRBs, being $\Delta K_{ \rm sd} $ much larger than 
$10^{50-51}$~erg, i.e. the typical energy emitted in X-rays and in
$\gamma$-rays during GRB events. We also remark that since $\Delta
t_{\rm sd}$ is much larger than the typical duration of the prompt
phase of LGRBs, deconfinement does not spoil the nice description of
the prompt emission of LGRBs within the proto-magnetar model.

However, as discussed before, also the HS to QS transition is
characterized by a huge release of energy (table~\ref{tab:transition})
that in principle could manifest itself as a second transient. Hence
the time for deconfinement {\bf $\Delta t_{\rm sd }$} is indicative of the
delay between the prompt GRB emission due to the HS and the possible
flare or second prompt emission associated to quark deconfinement.
Let us summarize how the deconfinement process proceeds and how is the
energy released. As studied in \citet{Drago_Pagliara15a},
deconfinement can be described as a combustion
process which can be separated in two phases. The first phase is very
rapid due to turbulence and it converts the bulk of the star in a time
scale of few ms \citep{Drago:2005yj,Herzog:2011sn}. The second phase
is dominated by diffusion of strangeness and it is therefore much
slower, typically lasting a few tens of seconds
\citep{Drago_Pagliara15a}. 
The huge energy associated to deconfinement is released via thermal neutrinos whose
luminosity can similarly be divided into an initial peak associated
with the deconfinement of the bulk of the star and a lower
quasi-plateau emission associated with the burning of the external
layer of the star. The interaction of neutrinos with the material of
the crust of the star causes the ablation of baryons which plays a
crucial role in the proto-magnetar model of GRBs. A distinctive
feature of the formation of a QS is the rapid suppression of
the baryonic flux once the conversion front reaches the surface of the
star.  This poses an upper limit to the duration of an event of the
prompt emission if described within the proto-magnetar model and
if associated with the baryonic emission from the surface of a star
undergoing quark deconfinement.  However, also the lepton emission
from the surface of the bare QS
\citep{Usov:1997ff,Usov:2001sw,Page:2002bj} can be used to produce a
jet.

As an example we discuss the case of GRB 110709B  characterized by two sub-bursts of observed duration 
respectively  $\Delta t ^{(1)}\sim 80$~s and $\Delta t ^{(2)}\sim 300$~s, separated by a delay 
$\Delta t_{\rm delay}\sim 600$~s \citep{Zhang_Burrows+12a,Penacchioni_Ruffini+13a}.
Neither the cosmological redshift nor the energetics of the event are well determined and in
the following we adopt values compatible with the analysis by  \citet{Penacchioni_Ruffini+13a}, being in
agreement with the pulse-wise Amati relation \citep{Basak_Rao13a}.
In particular we assume an isotropic energy $E_{\rm iso}^{(1)}=2.6\times 10^{53}$~erg for the first burst
and $E_{\rm iso}^{(2)}= 4.4\times10^{52}$~erg for the second one at redshift $z=1$. 
We further assume a beaming correction of $\sim 10^{-2}$ in line with average jet opening 
angle of LGRB \citep{Guidorzi_Mundell+14a}. 
Requiring that during the bursting events (radiatively-efficient phases) the mass loading enhances the spin-down
by a factor  $3-4$ with respect to the force-free case \citep{Metzger_Giannios+11a}, 
we find that, assuming an initial magnetic field $B=2\times 10^{15}$~G, HSs with initial spin period in the range $1-1.6$~ms
and baryon masses in $M_0= 1.72-1.70 M_{\odot}$  have enough rotational energy to power the 
first event in a time $\Delta t ^{(1)}$, deconfine with a delay $\Delta t_{\rm delay}$  and originate a QS rotating 
fast enough to power also the second event on the required time $\Delta t ^{(2)}$.
Notice that the requirements of short delays and fast rotating QS select for configurations with high baryon mass.
On the contrary HSs with lower $M_0$ take longer to decay and form a slowly rotating QS that at most could  power
weak flaring events in the afterglows, on timescales of few hours (see table ~\ref{tab:denergy}),
similarly to what is seen in GRB 050916 \citep{Chincarini_Moretti+07a}.

Finally, let us clarify how deconfinement can be responsible of rather
long gamma emissions, as in the case of GRB 110709B, and also of the
prompt phase of short GRBs, as discussed in
\cite{Drago_Pagliara15b,Drago_Lavagno+15a}. 
The main differences are related
to the neutrinos flux and energy which determine the baryonic mass
ejection rate.  In the case of short GRBs, the neutrino flux is
significantly larger (of at least one order of magnitude) mainly
because of the larger mass of the forming QS.  Also the
neutrino energy is larger roughly by a factor of three.  This implies
that the phase during which the QS forms cannot produce a
large enough Lorentz factor in the case of short GRBs.  At variance, in the
case of LGRBs, the mass ejection rate is not too large and the prompt
emission can start almost immediately at the beginning of deconfinement.

Although our model requires further investigation for a detailed assessment, 
our simple analysis of the spin-down evolution of CSs provides us with  
a new characterization, in terms of energetics and timescales,
of the possible observational signatures associated with the two-families scenario.
In particular we have shown, with reference to double GRBs, how, once the EoS of HSs and QSs are chosen,
the associated phenomenology is constrained by the initial magnetic field strength, 
initial rotational period and the stellar mass. This implies that, in principle, new observations could 
validate the two-families hypothesis, gaining new insights in the physics of dense matter.


\end{document}